\documentclass[twocolumn,showpacs,superscriptaddress,amsmath,amssymb,prl,nofootinbib]{revtex4}

\usepackage[pdftex]{graphicx}
\usepackage{dcolumn}
\usepackage{bm}
\usepackage{color}
\usepackage{SIunits}

\newcommand{\ud}[1]{{#1^{\dagger}}}

\newcommand{\ket}[1]{\left| #1\right\rangle}

\begin{document}

\title{Cavity versus dot emission in strongly coupled quantum
  dots--cavity systems}

\author{F.~P.~Laussy}
\email{fabrice.laussy@gmail.com}
\affiliation{Walter Schottky Institut, Technische Universit\"at M\"unchen, Am Coulombwall 3, 85748 Garching, Germany}%
\author{A.~Laucht}
\affiliation{Walter Schottky Institut, Technische Universit\"at M\"unchen, Am Coulombwall 3, 85748 Garching, Germany}%
\author{E.~del Valle}
\affiliation{School of Physics and Astronomy, University of Southampton, SO17 1BJ}
\author{J.~J.~Finley}%
\affiliation{Walter Schottky Institut, Technische Universit\"at M\"unchen, Am Coulombwall 3, 85748 Garching, Germany}%
\author{J.~M.~Villas-B\^{o}as}
\affiliation{Walter Schottky Institut, Technische Universit\"at M\"unchen, Am Coulombwall 3, 85748 Garching, Germany}%
\affiliation{Instituto de F\'isica, Universidade Federal de Uberl\^andia, 38400-902 Uberl\^andia, MG, Brazil}

\date{\today}

\begin{abstract}
  We discuss the spectral lineshapes of $N$ quantum dots in strong
  coupling with the single mode of a microcavity. Nontrivial features
  are brought by detuning the emitters or probing the direct exciton
  emission spectrum. We describe dark states, quantum nonlinearities,
  emission dips and interferences and show how these various effects
  may coexist, giving rise to highly peculiar lineshapes.
\end{abstract}

\maketitle


\section{Introduction}

The coherent coupling of a semiconductor quantum dot (QD) exciton to
the optical mode of a microcavity has been intensely investigated
throughout the last years in cavity quantum electrodynamics (CQED)
experiments~\cite{yoshie04a, reithmaier04a, peter05a, reitzenstein06a,
  hennessy07a, englund07a, faraon08a, kistner08a, winger08a,
  laucht09a, laucht09b, munch09a, reitzenstein09a, ota09a,
  suffczynski09a, thon09a, nomura10a, kasprzak10a, laucht10a,
  dalacu10a} and theory~\cite{andreani99a, cui06a, laussy08a,
  inoue08a, naesby08a, auffeves08a, laussy09a, delvalle09a,
  yamaguchi09a, hughes09a, auffeves09a, averkiev09a, richter09a,
  vera09a, vera09c, tarel10a, kaer10a, hohenester10a, ritter10a,
  poddubny10a}. In some of these works, the experimental spectral
function of the strongly coupled QD--cavity system was directly
compared to a theoretical model~\cite{laussy08a, laucht09b, munch09a,
  laucht10a}, and the agreement is excellent.  It was assumed that
most of the light escapes the system via the radiation pattern of the
cavity mode, and the experimental spectra were compared to the
spectral function calculated from the cavity occupation. This
detection geometry is known in atomic cQED as ``end emission'' or
``forward emission''~\cite{yamamoto_book99a}. In atomic systems,
negligible light escapes the cavity through the cavity mode, that is
to say, the cavity photon lifetime is so long as to be considered
infinite. Light is then detected in the so-called ``side-emission'',
where the radiation pattern of the emitter is probed instead. With
microcavities, the situation is reversed: the cavity mode is measured
often with an emitter of a much longer lifetime. In the spontaneous
emission regime of a system in strong-coupling, this makes
measurements of the Rabi doublet in the photoluminescence more
difficult, unless some cavity feeding makes the quantum state of the
system photon-like, since changing the nature of the excitation is, in
this case, equivalent to changing the channel of
detection~\cite{laussy08a}. In the nonlinear regime, this also hinders
manifestations of the Jaynes--Cummings ladder.  All the transitions
between its rungs have the same intensity in the exciton emission. In
the cavity emission, however, the photon has two paths to be emitted,
one with the dot in its ground state, the other with the dot in its
excited states~\cite{delvalle09a}. These two paths interfere
destructively when the initial and final states are out of phase,
which is the case for two out of the four possible transitions in the
Jaynes--Cummings ladder. On the other hand, these two paths interfere
constructively when the initial and final states are in phase, or, up
to a photon, indistinguishable. In the dressed-state picture, the
cavity photon to be emitted decouples from the polaritons and carries
away little information from the coupled system, being more like a
cavity photon the higher the number of excitations. The dot photon, on
the other hand, does not decouple from the system, regardless the
number of excitations: the dot cannot de-excite without altering
fundamentally the state of the entire system. As a result, the dot
photon carries more information of the coupled system. Summarizing,
the dot is essentially a quantum emitter whereas the cavity is
essentially a classical emitter.

It is therefore interesting to detect directly the dot emission. The
ratio~$\mathcal{R}$ of dot-vs-cavity emitted photons depends on the
populations of the dot (resp.~cavity), $n_1$ (resp.~$n_a$) and their
rate of emission, $\gamma_1$ (resp.~$\gamma_a$):
\begin{equation}
  \label{eq:ThuOct21102936CEST2010}
  \mathcal{R}=\frac{n_1\gamma_1}{n_a\gamma_a}\,.
\end{equation}
This ratio is typically small since---apart from the fact that $0\le
n_1\le 1$ whereas $n_a$ is unbounded---in typical experiments,
$\gamma_1\ll\gamma_a$. One should therefore take advantage of the
detection geometry, to detect light in a solid angle where the cavity
does not emit.  In a photonic crystal, one could filter out the areas
of most intense cavity emission by selective collection of the
farfield emission. The gain is however negligible, the cavity emission
being reduced by a factor of about $2$ as compared to the dot
emission\footnote{We estimated the ratio of cavity to dot emission by
  comparing the farfield emission pattern of a L3 cavity mode obtained
  from FDTD simulations with the emission of an isotropic
  emitter. From assuming collection for different numerical apertures
  or blocking of these numerical apertures we could estimate a maximum
  inhibition of the cavity emission compared to the dot emission of a
  factor $\sim2.5$.} when one needs several order of magnitude
increase to compensate~$\mathcal{R}$. The direct dot emission is
therefore more easily accessible with pillars~\cite{reitzenstein10a},
where one can detect on the side of the structure, and where the
feasibility of such experiments has already been
demonstrated~\cite{arXiv_sanvitto06a}.

In this text, we shall leave away the practical question of detection
and discuss the differences that are observed when probing the
strong-coupling physics in the cavity and the dot emission.  As the
possibility of coupling strongly $N>1$ dots to the (one) microcavity
mode~\cite{yeoman98a, delvalle07b, vera09b, arXiv_delvalle10a,
  poddubny10a} is starting to emerge
experimentally~\cite{reitzenstein06a, gallardo10a, laucht10a}, we will
consider the general case of many emitters. We address both the linear
and nonlinear regimes, with incoherent and continuous pumping as the
scheme of excitation, which is a favoured way of probing the system
experimentally. The effects we will discuss are not limited to the
case of quantum dots in a microcavity.  Systems with many identical
quantum emitters as in atomic physics~\cite{thompson92a,walther06a},
superconducting qubits~\cite{wallraff04a,fink09a,arXiv_filipp10} or
colour centers in diamond~\cite{Park06,Santori10,Englund10} would also
display the phenomenology we report.

\section{Model}


The Hamiltonian for $N$ independent excitons (in different QDs)
coupled to a common cavity mode reads
\begin{equation}
  \label{eq:ThuOct21105418CEST2010}
  H=\sum_{j=1}^N\left[{\omega_j\ud{\sigma_j}\sigma_j}+ g_{j}(a^{\dagger}\sigma_{j}+\ud{\sigma_{j}}a)\right]+\omega_{a}a^{\dagger}a,
\end{equation}
where $\ud{\sigma_{j}}$, $\sigma_{j}$ are the pseudospin operators for
the excitonic two level systems consisting of ground state $|0\rangle$
and a single exciton $|X_{j}\rangle$ state of the
$j$th--QD. $\omega_{j}$ is the exciton frequency, $a^{\dagger}$ and
$a$ are the creation and destruction operators of photons in the
cavity mode with frequency $\omega_{a}$, and $g_{j}$ describes the
strength of the dipole coupling between cavity mode and exciton of the
$j$th--QD. The incoherent loss and gain (pumping) of the dot-cavity
system is included in a master equation of the Lindblad form
$\frac{d\rho}{dt}=-{i}[H,\rho]+\mathcal{L}(\rho)$, where:
\begin{eqnarray}
  \mathcal{L}(\rho) &=& \sum_{j=1}^N\Big[\frac{\gamma_{j}}{2}(2\sigma_{j}\rho\ud{\sigma_{j}}-\ud{\sigma_{j}}\sigma_{j}\rho-\rho\ud{\sigma_{j}}\sigma_{j}) \nonumber\\
  &&+\frac{P_{j}}{2}(2\ud{\sigma_{j}}\rho\sigma_{j}-\sigma_{j}\ud{\sigma_{j}}\rho-\rho\sigma_{j}\ud{\sigma_{j}})\Big]\nonumber\\
  &&+\frac{\gamma_{a}}{2}(2a\rho a^{\dagger}-a^{\dagger}a\rho-\rho a^{\dagger}a)\nonumber\\
  &&+\frac{P_{a}}{2}(2a^{\dagger}\rho a-aa^{\dagger}\rho-\rho aa^{\dagger})\,.
\end{eqnarray}

Here, $\gamma_{j}$ is the $j$th exciton decay rate, $P_{j}$ is the
rate at which excitons are created by a continuous wave pump laser in
the $j$th QD, $\gamma_a$ is the cavity loss and $P_a$ is the
incoherent pumping of the cavity.  Pumping of the cavity from
non-resonant QDs was observed and investigated by different
groups~\cite{press07a,hennessy07a,kaniber08b,winger09a,suffczynski09a,hohenester09a,Chauvin09,ota09a,laucht10b}. It
has been shown how the effective quantum state realized in the system
under the interplay of the two types of pumping (cavity and exciton)
affects a lot the lineshape~\cite{laussy08a,laussy09a,delvalle09a},
which, in the linear regime, is a counterpart of the different
channels of emission.  A pure dephasing rate of the exciton in the
$j$th--QD could be included to account for effects originating from
high excitation powers or high temperatures~\cite{laucht09b}, which
has a tendency to merge together the closely spaced peaks arising from
higher rungs emission~\cite{gonzaleztudela10b}, so we take it
negligible for simplicity.

\begin{figure}[t]
\includegraphics[width=0.78\columnwidth]{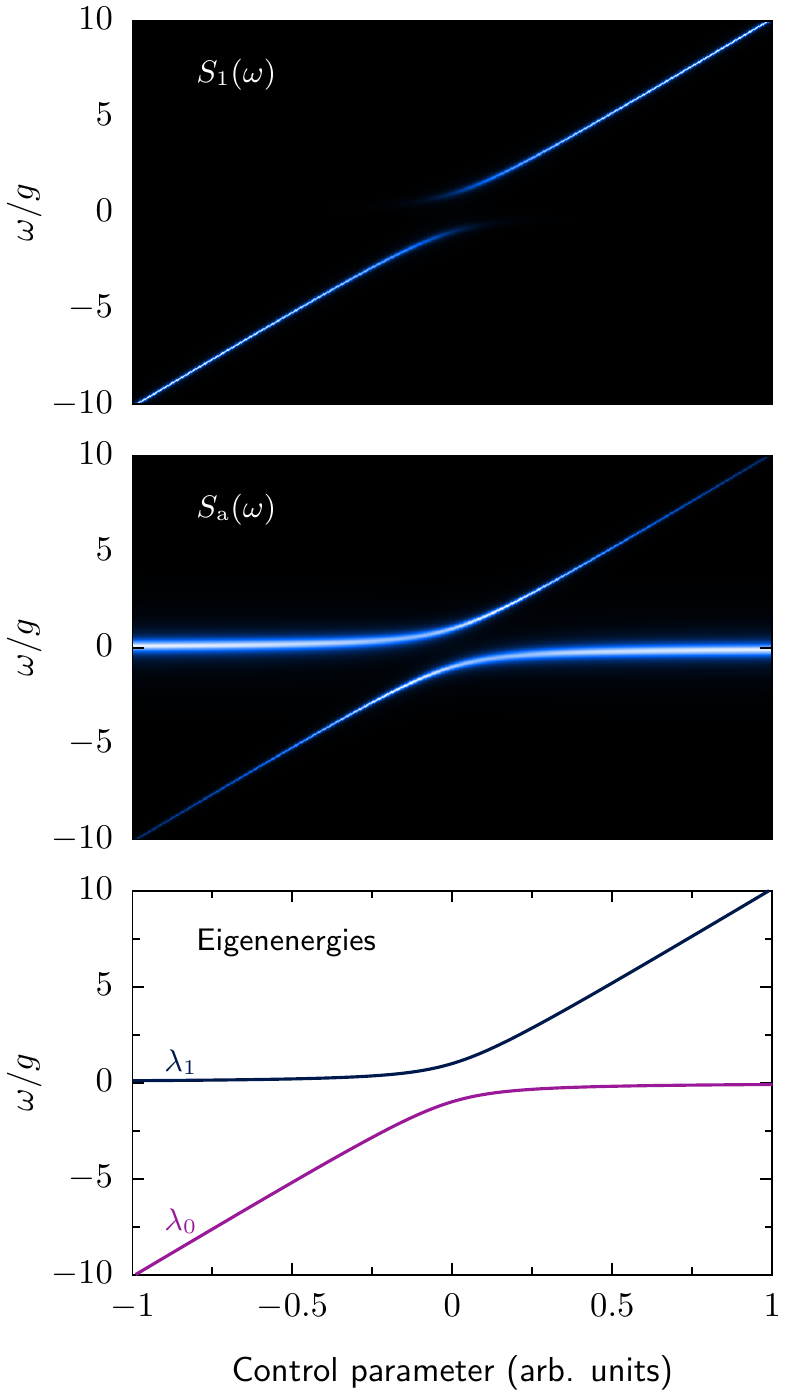}
\caption{\label{figure1} (Colour online) One dot in a cavity. Emission
  spectrum from the radiation channel of the exciton $S_1(\omega)$,
  from the radiation channel of the cavity mode $S_a(\omega)$, and
  eigenstates of the system. Parameters: $\gamma_a/g=0.5$,
  $\gamma_1/g=0.1$, $P_1/g=10^{-3}$ and~$P_a=P_1$.}
\end{figure}

\begin{figure*}[t!]
  \centering
  \includegraphics[width=\linewidth]{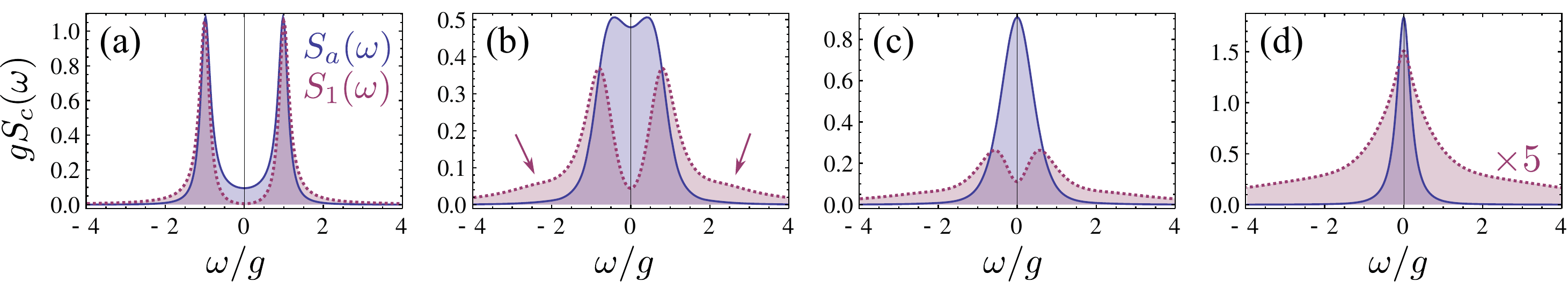}
  \caption{(Colour online) Cavity (solid blue) and dot (dotted purple)
    emission spectra for one dot in a cavity as a function of pumping
    power.  In the spontaneous emission regime (a), there are no
    qualitative differences between the two types of
    spectrum. Increasing pumping (b--d) shows markedly different
    behaviours in the two channels of emission.  Whereas the cavity
    spectrum does not present a particularly rich phenomenology
    (collapse of the Rabi doublet), the dot emission displays more
    characteristic lineshapes. Side elbows are formed [outlined with
    arrows in~(b)] and strong deviations from Lorentzian lines are
    obtained even when the spectrum has only one peak. Parameters:
    $\gamma_a/g=0.5$, $\gamma_1/g=0.1$, $P_a=0$ and $P_1/g=10^{-3}$,
    $0.5$, 1 and~2 from left to right panel.}
  \label{fig:ThuOct21180719CEST2010}
\end{figure*}

Assuming that the system achieves a steady state for long times and
employing the Wiener-Khintchine theorem, we calculate the spectral
function of our QD-exciton systems~\cite{Scully97} when photon
emission occurs via the normalized radiation pattern of the cavity
\begin{equation}\label{Scav}
S_a(\omega)=\frac{1}{n_a\pi}\lim_{t\rightarrow\infty}\mathrm{Re}\int_0^\infty d\tau e^{-(\Gamma_r-i\omega)\tau}\langle a^\dagger(t)a(t+\tau)\rangle,
\end{equation}
and via the normalized radiation pattern of the $j$th QD:
\begin{equation}\label{Sn}
S_{j}(\omega)=\frac{1}{n_j\pi}\lim_{t\rightarrow\infty}\mathrm{Re}\int_{0}^{\infty}d\tau e^{-(\Gamma_{r}-i\omega)\tau}\langle\ud{\sigma_j}(t)\sigma_{j}(t+\tau)\rangle.
\end{equation}
We included in the expression the term $\Gamma_r$ that takes into
account the finite spectral resolution of a
monochromator~\cite{Eberly77}, which is
$\approx\unit{18}\micro\electronvolt$ (half-width) for a good
monochromator by today standard. The qualitative effect of this term
is to broaden the peaks and blur the features. Therefore, in the
following we shall assume it is zero (case of a perfect detector).

Using the quantum regression theorem~\cite{Carmichael89}, the emission
eigenfrequency is obtained by solving the Liouvillian equations for
the single time expectation value, and has been amply detailed
elsewhere~\cite{delvalle_book10a}. Assuming that the emission energy
of different QDs change differently with respect to a control
parameter, which is the case for electrically tuning QDs, we can bring
two or more QDs at resonance at the same time. This can be simply
modeled by an effective control parameter such as
$\omega_j=\alpha_jV_\mathrm{control}$, where $\alpha_j$ gives the
different slopes of emission frequency with the control parameter
$V_\mathrm{control}$. The emission frequency of the cavity mode is
assumed not to be affected by this control parameter.

\section{One emitter}

We start with the simplest and most popular case of one QD in strong
coupling with the cavity mode.  We compare the emission spectra of the
cavity and dot emission in Fig.~\ref{figure1}, where we plot the
emission spectra from the radiation channel of the exciton
$S_1(\omega)$, from the radiation channel of the cavity mode
$S_a(\omega)$, and the eigenstates.  Close to resonance, both the
exciton and the cavity mode emit into both radiation channels and the
radiation patterns are very similar (see also
Fig.~\ref{fig:ThuOct21180719CEST2010}(a)). Far away from resonance,
the spectra look different, the excitonic channel $S_1(\omega)$ being
dominated by the emission of the exciton and the cavity channel
$S_a(\omega)$ being dominated by the cavity emission.  However, a
difference in the relative emission strengths cannot be directly
attributed to a preference in the radiation channel. It is also
dependent on the experimental parameters, on the specific pumping rate
of the exciton, the number of other transitions or QDs feeding the
cavity mode, and of course on the different coupling terms. In the
solid state environment the exciton is not only directly coupled to
the cavity mode, but also via the phonon
bath~\cite{hohenester09a,hohenester10a,kaer10a,arXiv_hughes11a}.

Increasing pumping, one reaches the nonlinear regime and climbs the
Jaynes--Cummings ladder~\cite{delvalle09a}. In the cavity emission, at
resonance, in a typical system where the coupling strength is not much
larger than the decay rates, this transition results in an apparent
collapse of the Rabi doublet into a single line. If the number of
photons is high enough, this line narrows as a consequence of the
cavity entering the lasing regime. This transition, displayed in solid
blue in Fig.~\ref{fig:ThuOct21180719CEST2010}, has been reported
experimentally~\cite{nomura10a}. Its counterpart in the dot emission
(purple dotted line) is richer in qualitative features in the
nonlinear regime~\cite{delvalle09a}. This manifests by the elbows in
Fig.~\ref{fig:ThuOct21180719CEST2010}(b-d) [indicated by arrows in
(b)], that arise from the transitions
$\ket{n,\pm}\rightarrow\ket{n-1,\mp}$, that are suppressed in the
cavity emission. We have used the standard notation for the
eigenstates of the Jaynes--Cummings Hamiltonian with~$\ket{n,\pm}$ the
state with $n$ excitation of higher ($+$) and lower ($-$) energy. In
the lasing regime, (d), the dot emission is strongly non-Lorentzian
and exhibits a characteristic lineshape reminiscent of the one-atom
laser~\cite{delvalle10d}.

\begin{figure}
  \centering
  \includegraphics[width=\columnwidth]{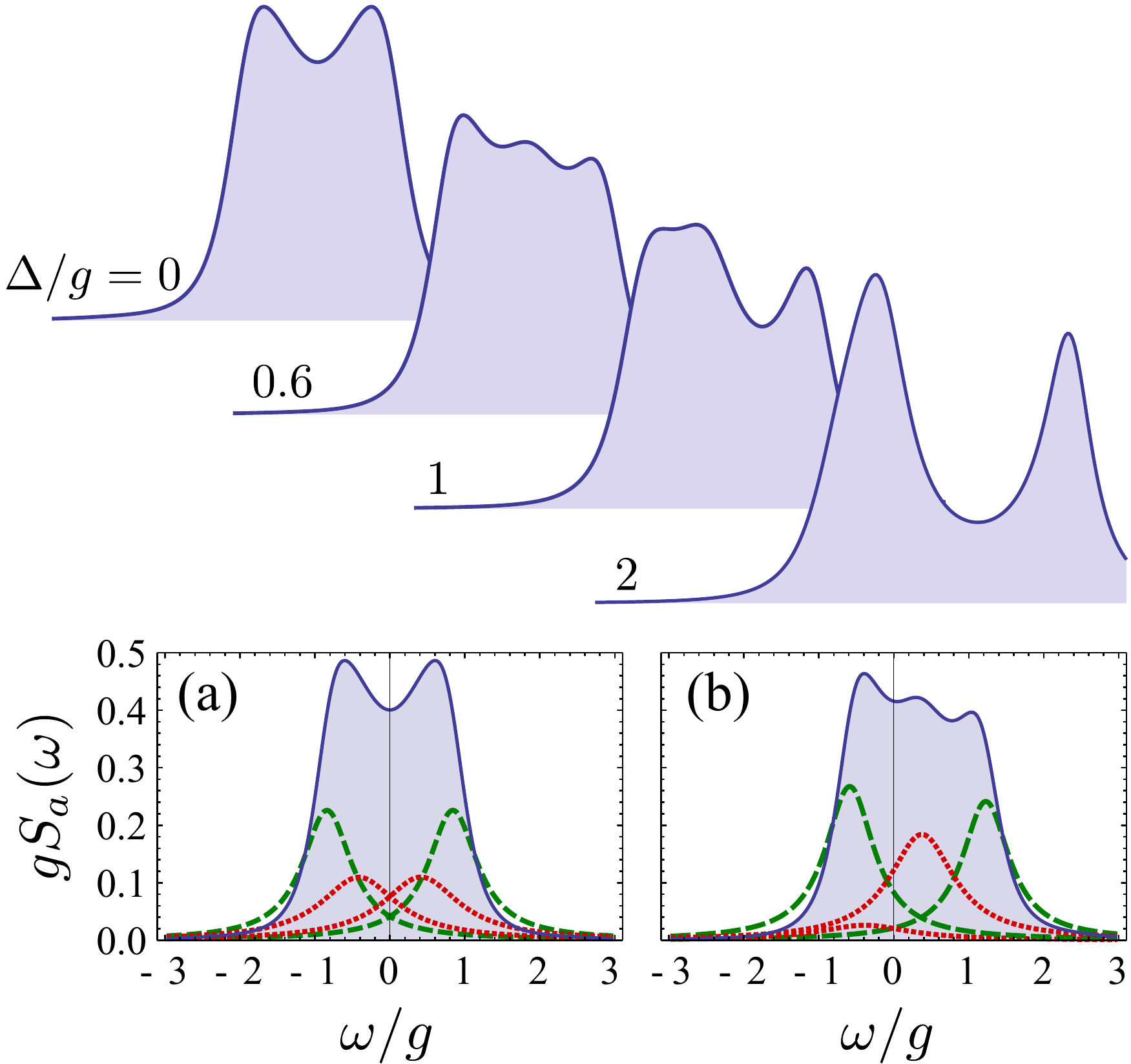}
  \caption{(Colour online) Cavity spectrum of one dot in a cavity as a
    function of detuning. Lower panels, decomposition of the spectrum
    in its dressed states emission lines
    ($\ket{1,\pm}\rightarrow\ket{\mathrm{vacuum}}$ in dashed green and
    $\ket{2,\pm}\rightarrow\ket{1,\pm}$ in dotted red), showing how
    (a) Jaynes--Cummings transitions are hindered at resonances but
    (b) are revealed out of resonance. Parameters are the same as in
    Fig.~\ref{fig:ThuOct21180719CEST2010} but with the detunings
    $\Delta/g=0$, $0.6$, $1$ and~$2$ and $P_1/g=0.4$.}
  \label{fig:ThuOct21190305CEST2010}
\end{figure}

A simple and convenient way to evidence Jaynes--Cummings
nonlinearities is to bring the system out of resonance. In this way,
transitions that are otherwise closely packed together, can be put
apart and resolved in the photoluminescence spectrum. This is shown in
Fig.~\ref{fig:ThuOct21190305CEST2010} for the same system as
previously but detuning the two bare emitters from each other by
$\Delta=\omega_a-\omega_1$. The Rabi doublet at resonance turns into a
triplet at small detunings. This is because one of the transitions
$\ket{2,\pm}\rightarrow\ket{1,\pm}$---that are too broad and too close
from the linear transitions
$\ket{1,\pm}\rightarrow\ket{\text{vacuum}}$ at resonance---can be set
apart at a small detuning, as shown in the lower panel~(a-b).

\begin{figure*}
  \centering
  \includegraphics[width=\linewidth]{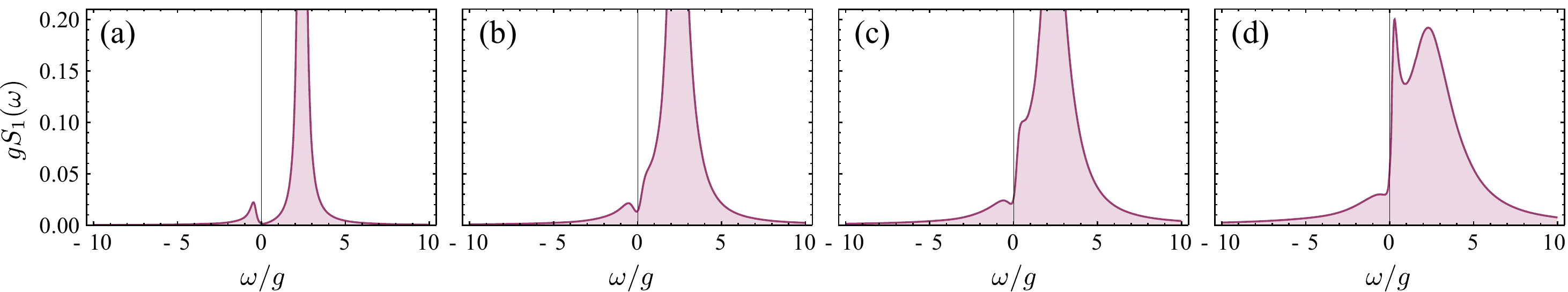}
  \caption{(Colour online) Dot emission spectrum of one dot in a
    cavity, out of resonance, as a function of pumping. The Rabi
    doublet in the spontaneous emission regime (a) gives rise, with
    increasing pumping (b--d), to an emission dip: the incoherent
    excitation is coherently scattered to the cavity which enters the
    lasing regime. This is better seen at nonzero detuning. Parameters
    are the same as in Fig.~\ref{fig:ThuOct21180719CEST2010} but for
    $\Delta=2g$.}
  \label{fig:ThuOct21200105CEST2010}
\end{figure*}

Transitions between dressed states provide a faithful mapping to the
exact system dynamics when the system is in very strong coupling, so
that dressed states are well defined and do not overlap appreciably
with each other. In the case where they overlap, say because the
splitting between dressed states is small or because their broadening
is large, interferences between the states enter the picture. The
luminescence can indeed be decomposed as a sum of Lorentzian emissions
from the dressed states with dispersive corrections arising from each
dressed states driving or being driven by the
others~\cite{delvalle09a}.  These interferences can become
particularly strong and complex when the system is brought towards the
classical regime, that is, with a lot of excitations. In this case,
many dressed states enter the collective dynamics, and their overlap
as well as mutual disturbance are thus much stronger.

This effect is also better seen at nonzero detuning.  Although the
interference grows with pumping also at resonance, it tends to be
cancelled by symmetry: dressed states from both sides of the origin
(set at the bare cavity emission) equilibrate each other. With
detuning, however, imbalance magnifies the interferences, that may
manifest strikingly. This is shown for instance in
Fig.~\ref{fig:ThuOct21200105CEST2010} that reproduces
Fig.~\ref{fig:ThuOct21180719CEST2010} (for the dot emission only) but
at detuning~$\Delta=2g$. The figure shows how, as pumping is
increased, an emission dip develops at the origin as the lines
broaden. It starts to be particularly visible in panel~(b) whereas, at
low excitation [panel~(a)], one merely sees the exciton, detuned,
favouring its own mode of radiation, just as in the linear case
(cf.~Fig.~\ref{figure1}). Note how, at resonance, in
Fig.~\ref{fig:ThuOct21180719CEST2010}, this interference is masked,
being essentially cancelled by symmetry.

\begin{figure}
\includegraphics[width=.9\columnwidth]{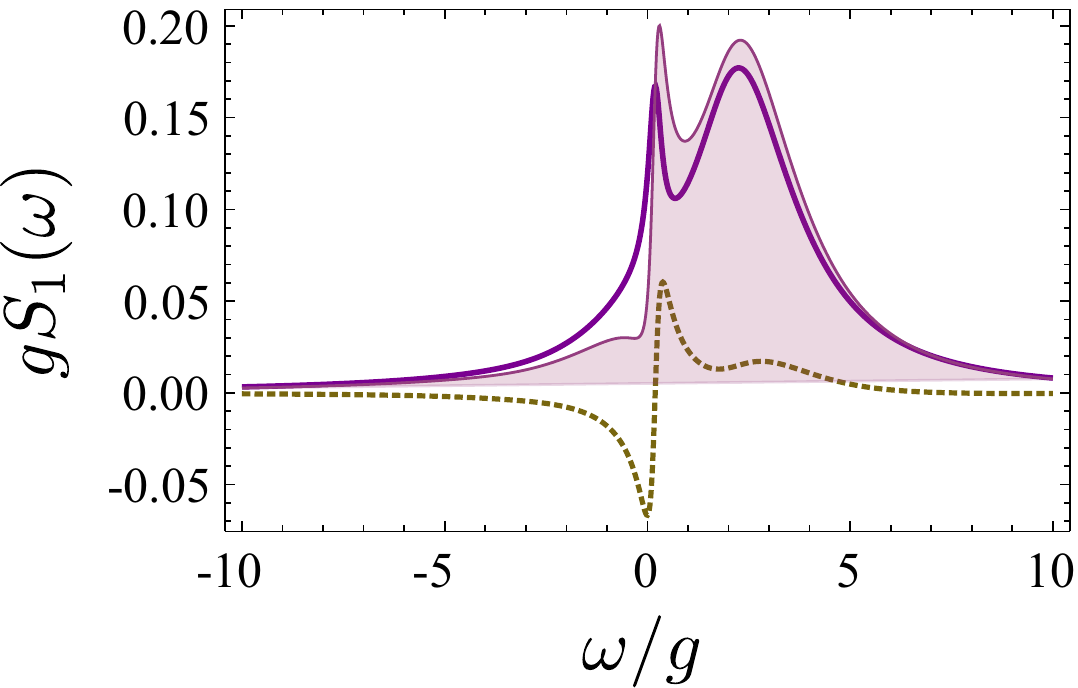}
\caption{(Colour online) Detail of
  Fig.~\ref{fig:ThuOct21200105CEST2010}(d). The interference that
  arises as the system enters lasing is best seen in the detuned
  system because dressed states (solid purple) are then of a markedly
  different character, cf.~the exciton line broadened by pumping,
  right, and the narrow line of the cavity that enters lasing
  (origin). The interferences between dressed states (dotted brown),
  which added to the dressed state emission provide the PL spectrum
  (filled purple), strongly modify the results of kinetic theory
  applied to the dressed states.}
\label{fig:FriJan14143603CET2011}
\end{figure}

The physical meaning of this dip can be assigned, in this case, to the
cavity entering the lasing regime at this frequency, and therefore
sucking-up excitations from the quantum dot, that is the medium
providing them. In Fig.~\ref{fig:FriJan14143603CET2011}, we show a
decomposition of Fig.~\ref{fig:ThuOct21200105CEST2010}(d) into the sum
of dressed states emission (thick solid purple) and the sum of
interferences between the dressed states (dashed brown). The observed
PL spectrum is the sum of these two contributions, obtained from a
mathematical decomposition of the
$G^{(1)}(t,\tau)=\langle\ud{a}(t)a(t+\tau)\rangle$ into terms that
give rise to Lorentzian lines (identifying the dressed states) and to
dispersive interferences when the dressed state emissions
overlap~\cite{delvalle09a}. The final result can be adequately
described by the dressed states only whenever interferences between
them are negligible, that is, when they do not overlap. This is the
case at low excitation and in the very strong coupling regime, when
$g\gg\gamma_a,\gamma_1$ and the splitting to broadening ratio of all
transitions is large. In this case, a kinetic theory that computes
mean occupation of the dressed states with rate (Boltzmann) equations
is adequate~\cite{poddubny10a}. Otherwise, a full master equation
approach is required, although it quickly becomes untractable. As the
intensity is further increased, interferences result in the breakdown
of the dressed state picture when the system acquires some macroscopic
coherence. Analytical investigations show that the emission dip
corresponds to a coherent scattering peak from the dot to the cavity,
which, in some approximations, becomes a Dirac $\delta$ function, that
describes Rayleigh scattering~\cite{delvalle10d}.

\section{More than one emitter}

The dynamics of strongly-coupled and strongly dissipative systems
becomes very complex as new paths of coherence flow between the
dressed states are opened by pumping and
decay. This can give rises to new peaks (of
emission or absorption) not accounted for by the dressed state
picture~\cite{delvalle10b}.

\begin{figure*}
\includegraphics[width=0.78\textwidth]{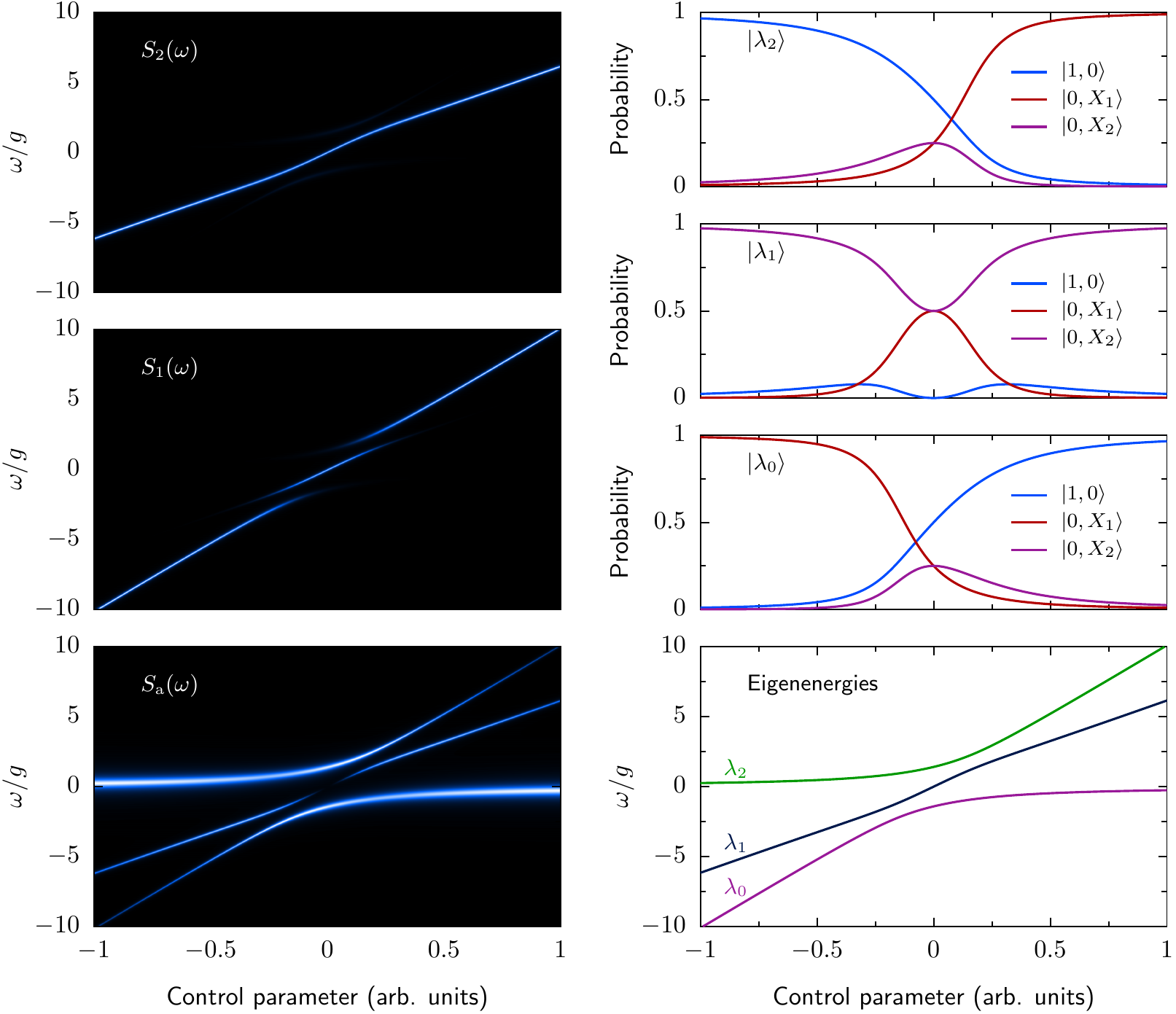}
\caption{\label{figure2} (Colour online) Emission spectrum from the
  radiation channel of the first exciton $S_1(\omega)$, the second
  exciton $S_2(\omega)$, the cavity mode $S_a(\omega)$, and the
  eigenstates for a strongly coupled system of two excitons and the
  cavity mode. The eigenvectors of the three eigenstates
  $\left|\lambda_2\right\rangle$, $\left|\lambda_1\right\rangle$, and
  $\left|\lambda_0\right\rangle$ of the coupled system display the
  contributions of the individual quantum states
  $\left|1,0\right\rangle$, $\left|0,X_1\right\rangle$, and
  $\left|0,X_2\right\rangle$ to the specific eigenstate. Parameters: same
  as in Fig.~\ref{figure1} with identical pumping of the dots.}
\end{figure*}

At low pumping, when all $N$-QDs excitons are exactly at resonance
with the cavity mode, the eigenvalues and eigenstates of the system
(neglecting incoherent loss, and setting the origin of the energy
scale to $\omega_a$) have a simple and well known
form~\cite{mandel_book95a}: two are splitted in energy,
$\lambda_{\substack{N\\0}}=\pm\Omega$ where
$\Omega=\sqrt{\sum_{j=1}^{N} g_j^{2}}$, while the other $N-1$ are
degenerated and equal to $0$. The corresponding eigenstates are:
\begin{subequations}
\begin{eqnarray}
  &&|\lambda_{\substack{N\\0}}\rangle=\frac{1}{\sqrt{2}}\left(\sum_{j=1}^{N}\frac{g_j}{\Omega}|0,X_j\rangle \pm|1,0\rangle\right)\,, \label{t6} \\
  &&|\lambda_{j-1}\rangle=\frac{1}{\Omega_j}\big(g_1|0,X_j\rangle-g_j|1,0\rangle \big)\,,\label{eq:FriFeb18165235CET2011}
\end{eqnarray}
\end{subequations}%
with $\Omega_j=\sqrt{g_1^2+g_j^2}$, with $j$ ranging from $2$ to $N$
for the degenerated eigenstates~(\ref{eq:FriFeb18165235CET2011}). From
this solution we can see that only the
$\ket{\lambda_{\substack{N\\0}}}$ states have a contribution from the
cavity photons. The cavity mode does not contribute to the other
states~$\ket{\lambda_{j-1}}$ which are called, for this reason,
\emph{dark states}. They consequently cannot be probed in the cavity
spectrum $S_a(\omega)$, but they can be very well seen in the
excitonic radiation channel or a mixture of all radiation
channels. These superpositions also give rise to the phenomenon of
sub- and superradiance, first reported by Dicke~\cite{dicke54a}. This
is essentially a classical effect, that is also observed with
vibrating strings. Recently, such configurations have been analyzed in
the microcavity QED context by Temnov and Woggon~\cite{temnov09a}, who
studied the photon statistics, and by Poddubny~\emph{et
  al.}~\cite{poddubny10a}, who studied the PL lineshapes. The latter
authors found that this classical regime is particularly fragile to
incoherent pumping since dark states, being also excited by pumping,
act as a long-lived reservoir for bright states from higher
manifolds. They also analyzed the regime of very high excitations,
when the system is (or is going towards) lasing. They observe in such
a case that, due to the predominance of the Dicke states that are the
most highly degenerated, the cavity spectrum is either oddly or evenly
peaked depending on the parity of the number of strongly coupled dots,
which is a strong manifestation in a readily measured observable of
the underlying microscopic configuration. In the following, we address
particular cases of larger than one, but still small number (given the
difficulty in coupling larger assemblies) of dots and show how, in the
linear regime, photoluminescence spectra vary greatly in a qualitative
way because of the contribution or suppression of the dark states. We
confirm that these states are quickly spoiled with increasing
pumping~\cite{poddubny10a} and give rise to nonlinear quantum
features, that also manifest in strikingly different ways depending on
the radiation channel that is probed.

\subsection{Two emitters}

In the case of two emitters coupled to a single cavity mode, the
eigenfrequencies in the linear regime can be obtained by solving the
eigenvalue problem given by the following equation:
\begin{equation}\label{eq2}
    i\frac{\partial}{\partial t} \begin{pmatrix}
    \langle a \rangle\\
    \langle \sigma_-^1 \rangle\\
    \langle \sigma_-^2 \rangle
    \end{pmatrix}=\begin{pmatrix}
    \tilde\omega_a & g_1 & g_2\\
    g_1 & \tilde\omega_1 & 0\\
    g_2 & 0 & \tilde\omega_2
    \end{pmatrix}
    \begin{pmatrix}
    \langle a \rangle\\
    \langle \sigma_-^1 \rangle\\
    \langle \sigma_-^2 \rangle
    \end{pmatrix}
\end{equation}
where $\tilde\omega_a=\omega_a-i\Gamma_a/2$ and
$\tilde\omega_j=\omega_j-i\Gamma_j/2$, with $\Gamma_a=\gamma_a-P_a$, and
$\Gamma_j=\gamma_j+P_j$. From the eigenstate of the
emission eigenfrequency we can obtain the degree of mixture of each
peak in the spectrum, i.e., the strength of the contribution of the
cavity mode, QD1 exciton and QD2 exciton to each individual
eigenstate.

In Fig.~\ref{figure2}, we investigate a system of two excitons in
different QDs simultaneously coupled to one cavity mode. We compare
the emission spectra obtained via the radiation channel of the first
exciton $S_1(\omega)$, the second exciton $S_2(\omega)$ and the cavity
mode $S_a(\omega)$. In the spontaneous emission regime, while all
three radiation channels exhibit a markedly different emission
spectrum, the most striking difference can be found in $S_a(\omega)$
where one of the emission lines vanishes completely. This occurs when
subradiance sets in, and follows for the case of two emitters an
analysis of the eigenvectors similar to that of
Ref.~\cite{laucht10a}. The plot of the eigenvectors in
Fig.~\ref{figure2} presents the contributions of the three quantum
states $\left|1,0\right\rangle$, $\left|0,X_1\right\rangle$, and
$\left|0,X_2\right\rangle$ to the three eigenstates
$\left|\lambda_2\right\rangle$, $\left|\lambda_1\right\rangle$, and
$\left|\lambda_0\right\rangle$ (as marked in the plot of the
eigenstates) of the coupled system. While
$\left|\lambda_2\right\rangle$ and $\left|\lambda_0\right\rangle$ have
contributions from all three quantum states at resonance, for the
eigenstate $\left|\lambda_1\right\rangle$ the contribution from
$\left|1,0\right\rangle$ goes to zero due to destructive
interference.

\begin{figure}
\includegraphics[width=0.78\columnwidth]{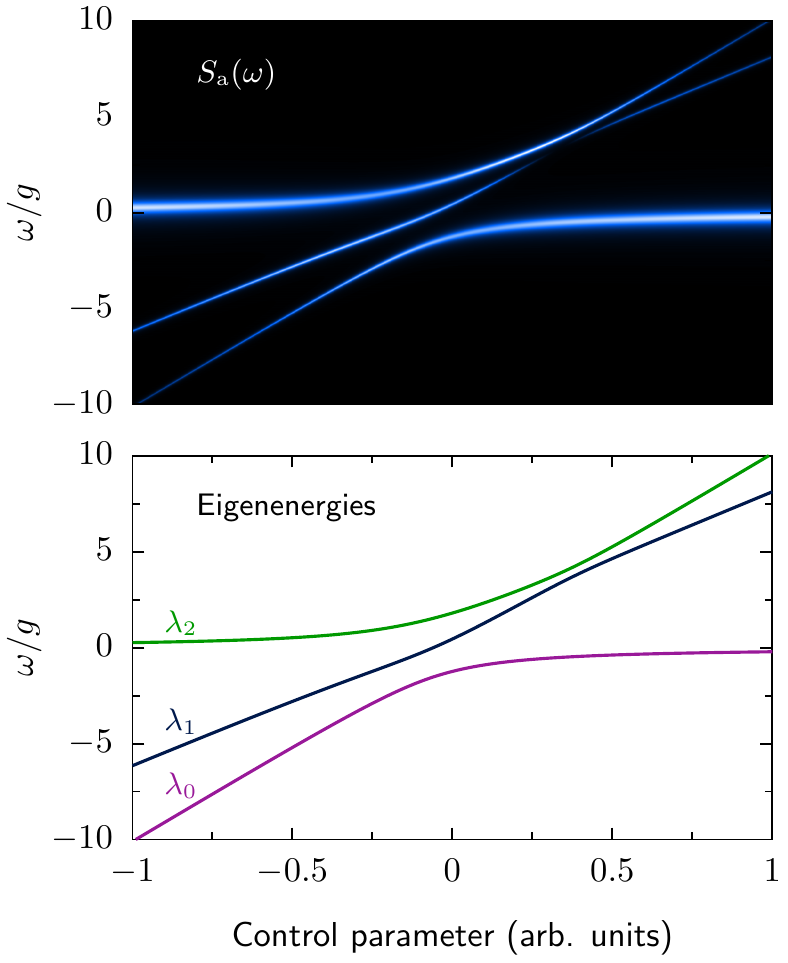}
\caption{\label{figure3} (Colour online) Emission spectrum from the
  radiation channel of the cavity mode $S_a(\omega)$, and eigenstates
  for a strongly coupled system of two excitons and the cavity mode
  where the two excitons cross out of resonance from the cavity
  mode. Parameters: same as in Fig.~\ref{figure2}.}
\end{figure}

\begin{figure}
  \includegraphics[width=.7\linewidth]{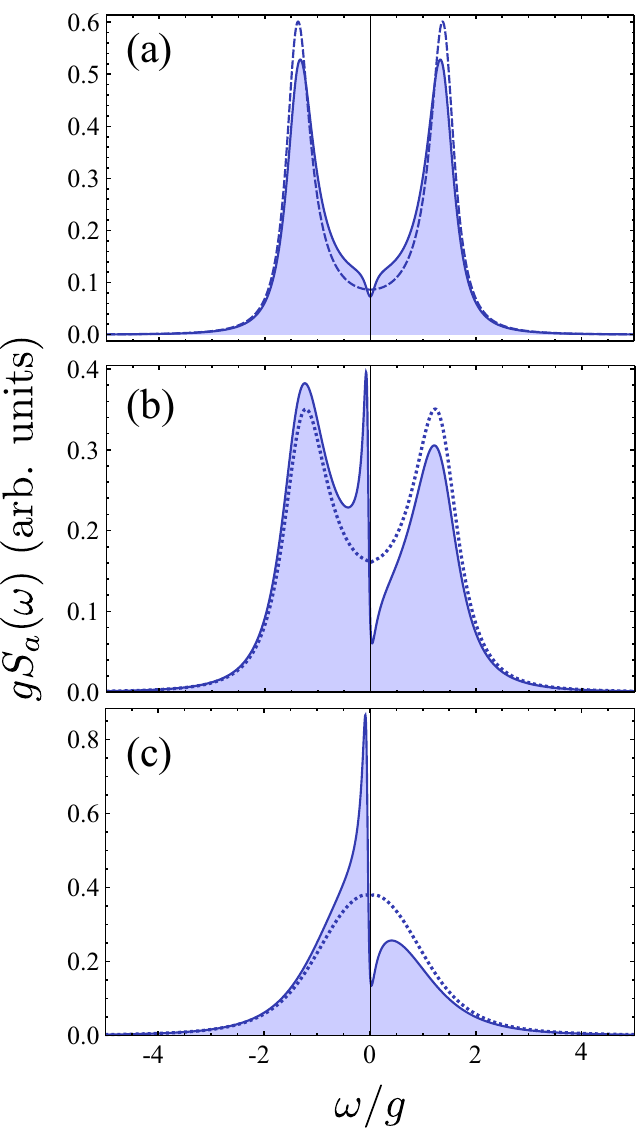}
  \caption{Spectral shapes in the cavity emission when two dots are
    strongly coupled to the cavity mode. (a) With increasing
    excitation (solid) the emission dip effect becomes visible in the
    cavity emission.  (b) It is better seen slighly out resonance
    since it can now ``feed'' on the line previously cancelled by
    subradiance. (c) Such interferences are strong even in systems
    that do not exhibit spontaneous emission features (such as Rabi
    splitting), altough this still requires
    strong-coupling. Parameters: $g_1=g_2=g$ (setting the unit),
    $\gamma_a=g$, $\gamma_1=\gamma_2=0.1g$, $P_a=0$, then: (a)
    $\Delta_1=\Delta_2=0$, $P_1=10^{-4}g$ with (dotted) $P_2=10^{-4}g$
    and (solid) $P_2=0.1g$. (b) Same as~(a) but for $\gamma_a=2g$,
    $P_1=10^{-3}g$, $P_2=0$, $\Delta_1=0$ and (solid) $\Delta_2=0.1g$
    or (dashed) $\Delta_2=0$.  (c) Same as~(a) but for $\gamma_a=5g$,
    $\Delta_1=0$ with (dotted) $\Delta_2=0$ or (solid)
    $\Delta_2=0.1g$.}
  \label{fig:MonOct25180806CEST2010}
\end{figure}

A measurement of this kind would also be possible for the same system
as just described, but when the two exciton lines anticross out of
resonance from the cavity mode, similar to the system described in
Ref.~\cite{laucht10a}. In Fig.~\ref{figure3}, we plot the emission
spectrum from the radiation channel of the cavity mode $S_a(\omega)$,
and the eigenstates of such a system. Probing the cavity emission,
one of the emission lines vanishes, similar to the case in
Fig.~\ref{figure2}, but this time when the two excitons are crossing
out of resonance from the cavity mode. In this situation the
eigenvalues and eigenstates have also a simple form,
$\lambda_{1}=\Delta$, $\lambda_{\substack{2\\0}}=\Delta/2 \pm
\sqrt{\Delta^2+4(g_1^{2}+g_2^{2})}/2$, with $\Delta=\omega_1-\omega_a$
being the mutual detuning from the cavity mode. In this case, the
eigenstates are:
\begin{eqnarray}
  &&|\lambda_{\substack{2\\0}}\rangle=\frac{1}{\sqrt{(\lambda_{\substack{2\\0}})^2+\Omega_2^2}}\big(g_1|0,X_1\rangle+g_2|0,X_2\rangle - \lambda_{\substack{2\\0}}|\mathrm{1,0}\rangle\big)\,,\nonumber \\
  &&|\lambda_{1}\rangle=\frac{1}{\Omega_2}\big(g_1|0,X_2\rangle-g_2|0,X_1\rangle \big)\,.
\end{eqnarray}
Again, since the contribution from the cavity mode to this particular
eigenstate ($|\lambda_{1}\rangle$) goes to zero, it cannot be probed
by the cavity radiation. The plot of all three eigenstates, however,
clearly shows the anticrossing behaviour of the two exciton-like
states when they come in resonance detuned from the cavity
mode.~\cite{laucht10a}. The disappearance of the second peak during
the anticrossing of the excitons brings a clear signature of a
collective strong coupling with the two dots and that this is probed
in the cavity radiation channel only.

This interference in the linear regime persists in the nonlinear
regime where it turns into the interference related to coherence
buildup in the system, as was the case with one dot in the cavity,
cf.~Figs.~\ref{fig:ThuOct21200105CEST2010}-\ref{fig:FriJan14143603CET2011}. Such
interferences are, however, now directly accessible through the cavity
spectrum, whereas they were previously only to be seen in the dot
emission, which is technically more challenging. This is shown in
Fig.~\ref{fig:MonOct25180806CEST2010}, at resonance~(a) and out of
resonance~(b-c). The sharp line near the origin in panels~(b) and~(c)
is the exciton line that appears suddenly as subradiance cancellation
is destroyed by going out of resonance, cf.~Fig.~\ref{figure2}.  It
results from the interplay of subradiance and detuning, studied in
Ref.~\cite{averkiev09a}, and the method outlined there, in the linear
regime, indeed reproduces such spectral features.  In the nonlinear
regime, this line also suffers from the dip carved by the cavity,
where it is sharply located. This effect is robust regardless of the
broadening of the cavity, i.e., with and without observation of the
Rabi doublet. Here again, detuning is paramount in revealing the
underlying physics, as seen in~(b-c) where the resonant case is
superimposed as dotted line: a very small dip at $\omega=0$ is hardly
visible at resonance [similar to the solid line in panel~(a)], in
contrast to the detuned case that produces a huge jump in the spectral
shape.

\subsection{Three emitters and beyond}


\begin{figure}[t!]
\includegraphics[width=0.8\columnwidth]{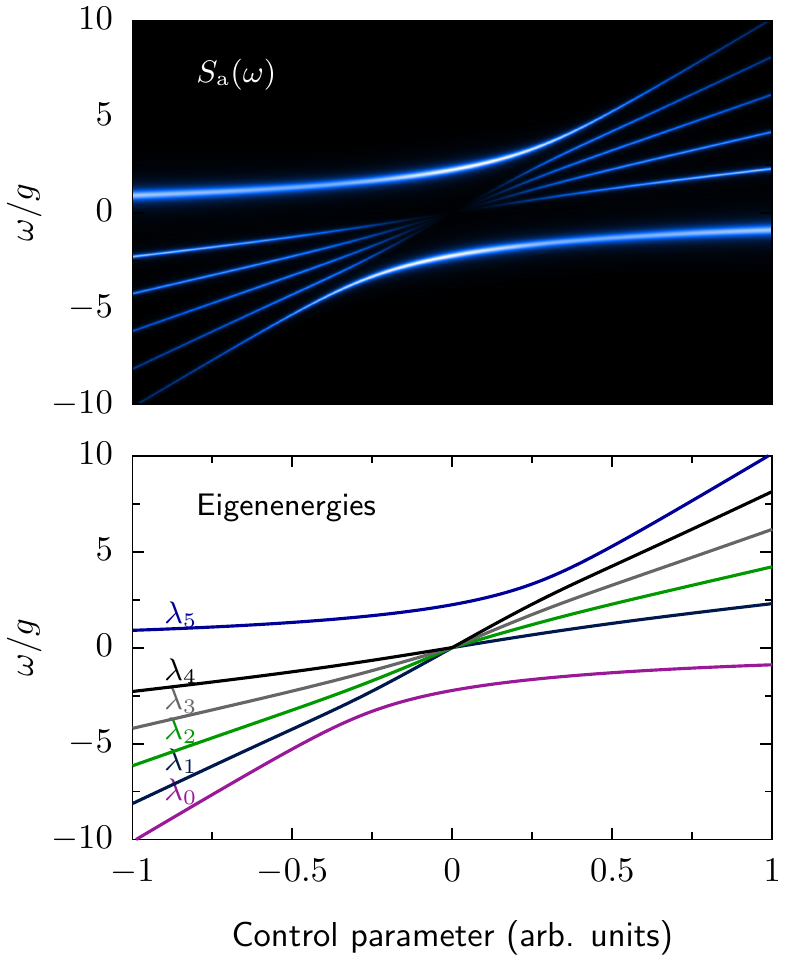}
\caption{\label{figure4} (Colour online) Emission spectrum from the
  radiation channel of the cavity mode $S_a(\omega)$ and the
  eigenstates of a strongly coupled system of five QD excitons and the
  cavity mode. Parameters: same as in Fig.~\ref{figure2}.}
\end{figure}

In the linear regime, the interference in the cavity emission due to
settling of dark states simply scales in the expected way with the
number of dots, by the very nature of linearity. As many lines vanish
as there are corresponding strongly coupled dots (minus one).
Exemplarily, we simulate the spectral shapes of a system of five
identical quantum emitters coupled to the same cavity mode and plot in
Fig.~\ref{figure4} the emission spectrum from the radiation channel of
the cavity mode $S_a(\omega)$ and the corresponding eigenvalues. At
resonance, the cavity radiation channel shows only the anticrossing of
two of the eigenstates of the coupled system, albeit with a larger
splitting corresponding to $2\Omega$, as described
before in Eq.~(\ref{t6}). 

\begin{figure}
  \includegraphics[width=.8\linewidth]{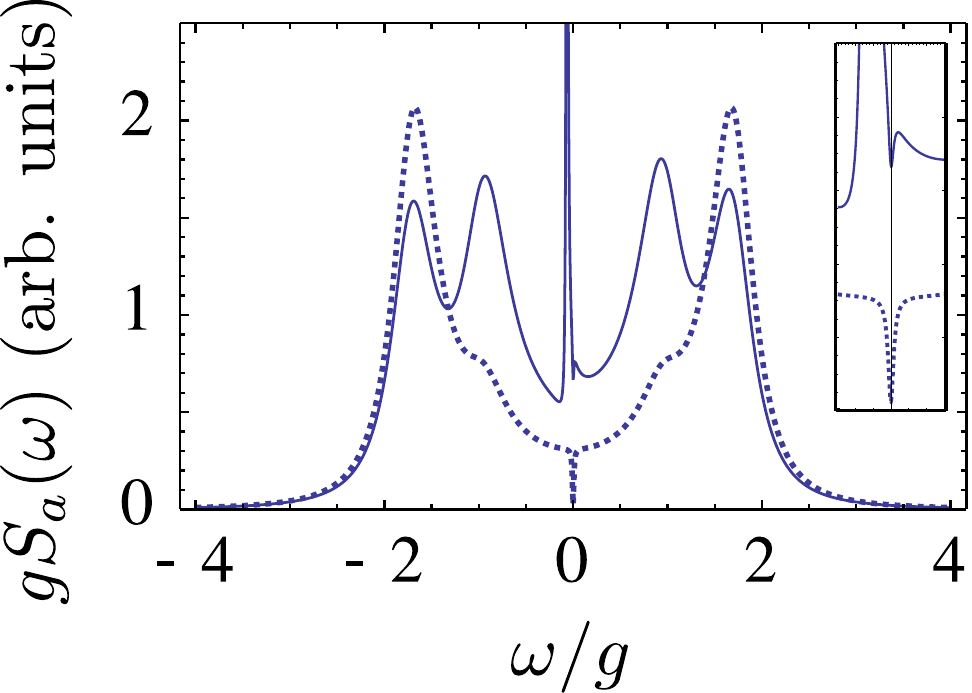}
  \caption{Spectral shapes in the cavity emission when three dots are
    strongly coupled to the cavity mode for low pumping (dashed line)
    and high pumping (solid line). This case brings together most
    features described above. At resonance (dotted), although the
    emission dip is seen at $\omega=0$, and also the peaks from higher
    rungs, these are faint and would be difficult to observe with
    finite detector resolution or counting noise. Detuning the system,
    (solid), destroys the dark state and gives rise to a sharp line
    nearby the origin. This one is then strongly affected by the
    emission dip. Finally, the peaks from nonlinear transitions become
    much more visible. Inset, zoom around the origin. Parameters:
    $g_1=g_2=g_3=1$, $\gamma_a/g=0.75$,
    $\gamma_1=\gamma_2=\gamma_3=0.004$, $P_1=P_2=10^{-6}g$,
    $P_3=0.01g$, $P_a=0$ and (dotted) $\Delta/g=0$ or (solid)
    $\Delta/g=0.1$.}
  \label{fig:MonOct25181639CEST2010}
\end{figure}

In the nonlinear regime, the situations discussed previously can be
reproduced and even brought together, with a higher degree of
complexity, by the very nature of nonlinearities.  A typical example
is shown in Fig.~\ref{fig:MonOct25181639CEST2010} for three emitters,
featuring the cavity emission at resonance (thick dotted) and slightly
out of resonance (thin solid). In both cases the emission dip is seen
at~$\omega=0$. In the former case, mainly the Rabi doublet, now
at~$\pm\sqrt{3}g$, is visible, with less clearly resolved peaks from
the transitions of multiply-excited states. Both these features would
be lost by finite spectral resolution of the detector. A small
detuning breaks the subradiance just like in the linear case and
results in the dot eigenstate showing up strongly as a very sharp
peak. This also displays the emission dip, like in the case with two
dots, as seen in the inset which is a zoom around the origin. This
case therefore brings together the two types of strong interferences
that manifest in this system: subradiance and emission dip. Note that
detuning, which unravels them, also makes much more prominent the
peaks that arise from quantum nonlinearities, producing a neat
quadruplet, a result we also put forward for the case of one emitter
in strong coupling with the cavity.

\section{Conclusions}

We investigated theoretically spectral shapes of $N$ QDs strongly
coupled to a single mode of a microcavity, and compared the emission
spectra obtained from the two different radiation channels offered by
the cavity and direct quantum dot emission, both in the linear and
nonlinear regime, at and out of resonance. In the spontaneous emission
regime, dark states forming at resonance result in a vanishing of
spectral lines in the cavity emission only.  When excitation is
increased and multiple-photon effects become important, dressed states
enter the picture. These are difficult to see in state-of-the-art
experiments where splitting to broadening ratio does not allow them to
be clearly resolved. Although strong-coupling is maximum at resonance,
we find that detuning is an agent to reveal the quantum nonlinear
features and that here too dot emission behaves qualitatively
differently from cavity emission. As excitation is further increased
and a large number of cavity photons is generated (the system enters
lasing), another interference due to onset of coherence takes
place. It manifests as an emission dip that results from coherent and
elastic scattering between the modes.  A rich phenomenology thus
remains to be observed in these systems. One can access it either by
detecting direct dot emission---which is more difficult
technically---or by considering strong-coupling involving more than
one emitter. In both cases, detuning is a powerful tool to unravel
this new physics, since, although optimum, strong-coupling is balanced
and/or cancelled at resonance.

We gratefully acknowledge comments from Dr.~Poddubny and Dr.~Glazov
and financial support of the DFG via the SFB 631, Teilprojekt B3, the
German Excellence Initiative via NIM, and the EU-FP7 via SOLID. JMVB
acknowledges the support of the Alexander von Humboldt Foundation,
CAPES, CNPQ and FAPEMIG, EdV of a Newton International fellowship and
FPL of the FP7-PEOPLE-2009-IEF project ‘SQOD.’

\bibliography{Papers,Sci,books,arXiv}

\end{document}